\documentstyle[prd,aps,epsfig]{revtex}

\def \Journal#1#2#3#4{{#1} {\bf #2}, #3 (#4)}

\def \NPB{{\em Nucl. Phys.} B}
\def \PLA{{\em Phys. Lett.}  A}
\def \PLB{{\em Phys. Lett.}  B}
\def \PRL{\em Phys. Rev. Lett.}

\def \PRD{{\em Phys. Rev.} D}

\def \GRG{\em Gen. Rel. Grav.}
\def \CQG{\em Class. Quantum Grav.}
\def \JPA{\em Journal of Physics A}
\def \JETP{\em Sov. Phys. JETP}

\begin{document}
\draft
\preprint{Accepted for publication in \prd}
\title{Order reduction in semiclassical cosmology.}
\author{Gra\.zyna Siemieniec-Ozi\c{e}b{\l}o and Andrzej Woszczyna}
\address{Astronomical Observatory, Jagellonian University,\\ ul. Orla 171, 30-244 Krak\'ow, Poland}
\address{woszcz@oa.uj.edu.pl}
%\author{C. C. Author}
%\address{}
%\date{\today}
\maketitle
\begin{abstract}
We investigate the Robertson-Walker cosmology with Lagrangian
$R+\alpha_1\hbar R^2+\alpha_2\hbar R^{\mu\nu}R_{\mu\nu} +L_{rad}$ where
$L_{rad}$ means classical source with traceless energy-momentum tensor.
We weaken the self-consistence condition (L. Parker, J. Z. Simon, Phys.
Rev. {D47}(1993),{1339}). Quantum corrections are expressed as
contributions to the effective equation of state. We show that the empty
space-time is stable within the class of radiation-filled expanding
universes with no order reduction of the field equations.

\end{abstract}
\pacs{PACS numbers: 9880}
%\pacs{Valid PACS appear here.
%{\tt$\backslash$\string pacs\{\}} should always be input,
%even if empty.}

\section{Introduction}
Controversial fourth order differential equations, which govern the
semiclassical cosmology can be reduced to second-order \cite{Simon},
\cite{PS93}, and in this way, exempted from quantum-originated
instabilities \cite{kwantinst} \cite{starobinski}. The reduction is
based on the {\it self-consistence} condition, i.e. the assumption that
both equations and solutions are perturbatively expandable in $\hbar$.
Under this condition the universe becomes an ordinary mechanical system
with a two-dimensional phase-space corresponding to the single degree of
mechanical freedom -- the scale factor $a(t,\mbox{$\hbar$})$.
Self-consistent theory is still renormalizable \cite{Mazzitelli},
Minkowski space-time regains stability in the class of homogeneous and
isotropic models, quasi-inflationary phenomena disappear \cite{Simon2}.
Similar reduction techniques are being applied to gravity with higher
than fourth-order derivatives \cite{Schmidt} and also in other branches
of physics \cite{Agu}.

However, imposing the self-consistence condition on the cosmological
scale $a(t)$  encounters some difficulties. In a universe with vanishing
spatial curvature still remains a freedom to multiply metrics by an
arbitrary constant factor, therefore the scale $a(t)$ is not a
measurable quantity. Requirement for $a(t,\hbar)$ to be
$\hbar$-expandable is physically unclear. In open or closed universes
this freedom is reduced by the choice $k=\pm 1$, and the scale factors
are uniquely determined by cosmological observables: the Hubble
parameter $H$  and the energy density $\epsilon$. Yet arbitrarily small
changes in any of these observables in the vicinity of critical density
$\epsilon -\Lambda =3H^2$ may result in indefinite changes of $a(t)$.
The perturbative character of the energy-momentum tensor (and consequently
the equations) with respect to an arbitrary chosen parameter, in
general, does not imply the same property of the metrics. Finally,
expanding $a(t)$ in the equations, which contain fixed curvature index
$k$ \cite{PS93}, limits quantum corrections to only those, which
preserve the same sign of the space curvature. This limitation is
particularly severe for a flat universe, where generic quantum corrections
would contribute to the space curvature unless the $k=0$ condition
prevents that. This limitation cannot be derived directly from the
Lagrangian and, in fact, it forms an additional constraint imposed on the
theory (which is not even true in classical gravity\footnote{Note
that the Lema{\i}tre universes, which are of positive space curvature,
are obtained from flat universes (not closed!) when $\Lambda$ diverges
from zero.}).

Not arguing with the very idea of self-consistence, we draw attention to
some circumstances which are important for semiclassical cosmology:

1. Without harm to the reduction procedure, one can release the
consistence condition for the scale factor, demanding instead the same
property for cosmological observables (the Hubble parameter, etc.)

2. For a radiation filled universe with vanishing cosmological constant
$\Lambda=0$ the self-consistence condition is superfluous, since the original
equation is of second (!) order. Terms with higher derivatives cited
by classical papers \cite{FHH} contain an additional hidden factor
$\hbar$ and are eventually eliminated in the first order expansion.

3. We show that quantum corrections form the equation of state of a
barotropic fluid, and discuss the stability of the Minkowski space-time
on the ground of dynamic systems theory.

\section{Condition of self-consistence for Hubble's expansion rate.}

We consider semiclassical gravity theory with the Lagrangian
$R+\alpha_1\hbar R^2+\alpha_2\hbar R^{\mu\nu}R_{\mu\nu} +L_{rad}$, where
$L_{rad}$ represents classical radiation or another thermalized field of
massless particles. Typically, cosmologies containing the $R^2$ and the
$R^{\mu\nu}R_{\mu\nu}$ terms lead to 4 order equations \cite{FHH} and
violate the stability of empty space\cite{kwantinst}.

We write quantum terms on the right-hand side of the field equations and
treat them as corrections to the energy-momentum tensor. We think of
$\hbar$ as the theory parameter, which can take arbitrary values, so
the limit transition $\hbar\rightarrow 0$ defines the classical limit of
the theory. The field equations we write in the Einsteinian form
$R_{\mu\nu}-\frac12Rg_{\mu\nu}+\Lambda g_{\mu\nu}=T_{\mu\nu}$, but with
the modified, effective energy-momentum tensor

\begin{equation}T_{\mu\nu}=T_{\mu\nu}^{(rad)}
-\hbar\alpha_1\phantom{*}^{(1)}H_{\mu\nu}
-\hbar\alpha_3\phantom{*}^{(3)}H_{\mu\nu},\label{Tef}\end{equation}

\noindent where

\begin{eqnarray}
\phantom{a}^{(1)}H_{\mu\nu}&=&\textstyle{\frac 1 2}R^2g_{\mu\nu}
-2RR_{\mu\nu}-2\Box Rg_{\mu\nu}+2\nabla_{\mu}\nabla_{\nu}R\\
\phantom{a}^{(3)}H_{\mu\nu}&=&-R_{\mu}^{\sigma}R_{\sigma\nu}
+\textstyle{\frac 23}RR_{\mu\nu}+\textstyle{\frac 12}R_{\sigma\rho}
R^{\sigma\rho}g_{\mu\nu}-\textstyle{\frac 1 4}R^2g_{\mu\nu}\end{eqnarray}

\noindent and the constant $\alpha_3$ is some combination of $\alpha_1$ and
$\alpha_2$ (Robertson-Walker symmetry have been partially exploited
to derive formula (\ref{Tef}). For more precise explanation see \cite{PS93}.)
Derived in this way the (0,0)-equation

\begin{eqnarray}
0&=&-\Lambda -\frac {\kappa\mu}{a^4}+\frac 1{a^2}
\left[3k+3\left(\frac {da}{dt}\right)^2\right]\nonumber\\
&+&\frac {\alpha_1\hbar}{a^4}\left[-18k^2+36k\left(\frac {da}{dt}\right)^2
+54\left(\frac {da}{dt}\right)^4-36a\left(\frac {da}{dt}\right)^2\frac {d^2a}{dt^2}
+18a^2\left(\frac {d^2a}{dt^2}\right)^2-36a^2\frac {da}{dt}\frac {da^3}{dt^3}\right]\nonumber\\
&+&\frac {\alpha_3\hbar}{a^4}\left[3\left(\frac {da}{dt}\right)^4+6k\left(\frac {da}{dt}\right)^2
+3k^2\right]\label{a1}\end{eqnarray}

\noindent contains four fundamental constants, two of them classical -
the gravitation constant $\kappa$ (further on we put $8\pi\kappa=1$),
the cosmological constant $\Lambda$, and two quantum ones - $\alpha_1$
and $\alpha_3$. There are also two quantities which define a particular
solution: the constant of motion $\mu =\epsilon_0a_0^4$, and the index
of space curvature $k$. Therefore, the transition from classical to
quantum theory with the self-consistence  $a=a_0+\hbar a_1$ imposed on
(\ref{a1}) preserves the type of space curvature, including the strong
$k=0$ limitation for the flat universe.

One can get rid of the last two constants, and consequently, of the
constraints they bring, by introducing the Hubble expansion parameter
$H=\frac 1a\frac {da}{dt}$. Differentiating (\ref{a1}) twice, we obtain
the fourth order equation for $H$, which contains only fundamental
constants\footnote{The reverse procedure would give the equation with
two parameters of continuous values. Consequently, the equation (\ref{4H})
formally has a broader class of solutions
than (\ref{a1}). However, the freedom to choose $k$ as
different from $0$, $\pm 1$ is a trivial one, and resolves itself to
rescaling the metrics by a constant factor.}
  
\begin{eqnarray}
0&=&-3\frac {d^2H}{dt^2}-18H\frac {dH}{dt}-4H\left(3H^2-\Lambda\right)\nonumber\\
&+&18\hbar\alpha_1\frac {d^4H}{dt^4}+162\hbar\alpha_1H\frac {d^3H}{dt^3}\nonumber\\
&+&\frac {d^2H}{dt^2}\left[6\left(51\hbar\alpha_1
+\hbar\alpha_3\right)\frac {dH}{dt}+6\left(90\hbar\alpha_1
+\hbar\alpha_3\right)H^2-4\Lambda\left(6\hbar\alpha_1+\hbar\alpha_3\right)\right]\nonumber\\
&+&4H\left[162\hbar\alpha_1\left(\frac {dH}{dt}\right)^2
+\frac {dH}{dt}\left(3\left(48\hbar\alpha_1-\hbar\alpha_3\right)H^2
-2\Lambda\left(6\hbar\alpha_1+\hbar\alpha_3\right)\right)
-3\hbar\alpha_3H^4\right].\label{4H}
\end{eqnarray}

\noindent This equation describes the dynamics of Robertson-Walker models with
arbitrary space curvature, and what is equally important, it is
expressed in terms of observable quantities. A self-consistence condition
imposed on measurable quantities has well defined physical meaning. We
adopt Simon's ansatz to $H$, namely we state that
$H(t)=H_{class}(t)+\hbar H_{quant}(t)$ is perturbative in $\hbar$. Now,
the procedure of the order reduction can be done in two ways:\\ 1) one
can differentiate twice the zeroth-order expansion (equation
(\ref{4H}) with $\alpha_1=\alpha_3=0$) to find the third and fourth derivatives and
eliminate them from the full equation (\ref{4H}) - this is equivalent to
what is done in \cite{PS93},\\ 2) substitute the expansion
$H(t)=H_{class}(t) +\hbar H_{quant}(t)$ directly into (\ref{4H}) and
abandon  terms second order in $\hbar$ or higher .\\ In both cases we obtain
the second order equation

\begin{eqnarray}
0&=&-3\frac {d^2H}{dt^2}-18H\frac {dH}{dt}-4H\left(3H^2-\Lambda\right)\nonumber\\
&+&2\frac {d^2H}{dt^2}\left[3\left(51\hbar\alpha_1+\hbar\alpha_3\right)
\frac {dH}{dt}+3\left(90\hbar\alpha_1+\hbar\alpha_3\right)H^2
-2\Lambda\left(6\hbar\alpha_1+\hbar\alpha_3\right)\right]\nonumber\\
&+&4 H\left[459\hbar\alpha_1\left(\frac {dH}{dt}\right)^2
+\frac {dH}{dt}\left(3\left(372\hbar\alpha_1-\hbar\alpha_3\right)H^2
-2\Lambda\left(69\hbar\alpha_1+\hbar\alpha_3\right)\right)\right]\nonumber\\
&+&4\left(3\left(180\hbar\alpha_1-\hbar\alpha_3\right)H^4
-204\Lambda\hbar\alpha_1H^2+8\Lambda^2\hbar\alpha_1\right),\label{2H}\end{eqnarray}

\noindent which is nonlinear both in $H$ and its derivatives. So
strong nonlinearity allows one to find exact solutions only in some
particular situations. The is not the case in equation (\ref{2H}).
However, this equation becomes much more transparent after one
rewrites the quantum corrections as contributions to energy density and
pressure. Qualitative analysis is then enabled.

Let $\epsilon$ and $P$ denote respectively effective energy density and
effective pressure, i.e. each of these quantities is supplemented by
quantum corrections. The universe dynamics is determined by the system
of the Raychaudhuri (\ref{Ray}) and the continuity (\ref{kont})
equations

\begin{eqnarray}
\frac {dH}{dt}&=&-H^2-\frac 16\left(3P+\epsilon -2\Lambda\right)\label{Ray}\\
\frac {d\epsilon}{dt}&=&-3H\left(P+\epsilon\right)\label{kont}\end{eqnarray}

\noindent We differentiate (\ref{Ray}), substitute into (\ref{2H}) and
apply the continuity equation (\ref{kont}) to get the relation between
pressure, energy and the cosmological constant in differential form

\begin{equation}
\frac {dP}{d\epsilon}=\frac {P+\epsilon /9}{P+\epsilon}
-\frac 29\alpha_3\hbar\frac {\left(3P+\epsilon\right)^2}
{P+\epsilon}-\frac {\alpha_3\hbar}{27}\frac {8\Lambda^2}
{P+\epsilon}
\label{diffstate}\end{equation}

\noindent As a matter of fact, one can solve equation
(\ref{diffstate}) analytically, however the solution takes unclear
implicit form. This is much simpler to follow the other way. The
solution of (\ref{diffstate}) must be a function of the energy density
and cosmological constant  solely, hence $P(\epsilon ,\Lambda )$ is
independent of the expansion rate $H$. Therefore the limit transition
$H^2\rightarrow 3(\epsilon +\Lambda )$ does not affect its values,
and the general solution is identical with the integral
found for the flat universe. In the last case the equation

\begin{equation}
-\frac 1{18}\left[3\frac {dH}{dt}+2\left(3H^2-\Lambda\right)\right]
+\hbar\alpha_1\frac {d^3H}{dt^3}+7\hbar\alpha_1H\frac {d^2H}{dt^2}
+\frac {\hbar}3\left[12\alpha_1\left[\frac {dH}{dt}\right]^2
+\left(36\alpha_1-\alpha_3\right)H^2\frac {dH}{dt}-\alpha_3H^4\right]=0
\end{equation}

\noindent is an analogue to equation (\ref{4H}). Its order reduces by
two, and finally the equation takes a particularly simple form

\begin{equation}\frac {dH}{dt}=-\frac 2 3\epsilon
+\frac{2\alpha_3\hbar}9\left(^{}\epsilon^2-\Lambda^2\right)
\label{redflat}\end{equation}

\noindent Now, comparing (\ref{redflat}) with the Raychaudhuri equation
(\ref{Ray}) we obtain the equation of state of cosmological substratum
in the form of the algebraic relation

\begin{equation}P=\frac 13\epsilon -\frac{4\alpha_3\hbar}9
\left(^{}\epsilon^2-\Lambda^2\right)
\label{algestate}\end{equation}

\noindent Function $P(\epsilon ,\Lambda )$, defined by
(\ref{algestate}), fulfills the differential equation(\ref{diffstate})
with an accuracy to terms $o(\hbar )$. By simple calculation
\cite{SWAC}, one can confirm that the exact solutions found by Parker
and Simon also obey (\ref{algestate}).

As we have already mentioned, the equation of state (\ref{algestate}) is
barotropic, i.e. effective pressure is solely the function of the
effective energy density (including the energy of vacuum $\Lambda$).
While reducing the equations order we eliminate contributions to the
energy-momentum tensor coming from the expansion rate \cite{FHH};
therefore the universe evolution becomes a reversible process (equations
(\ref{Ray})-(\ref{kont}) are invariant under the time reflection
$t\rightarrow -t$).

Quantum corrections contained in (\ref{algestate}), and consequently the
dynamical system (\ref{Ray}-\ref{kont}) are free of the $\alpha_1$
constant. The only term multiplied by $\alpha_1$ which survives the
reduction procedure \cite{PS93}, has been assimilated here by the
effective energy density\footnote{In this approach the quantum
corrections modify effective energy density and pressure, not the
fundamental constants like in \cite{PS93}}.

\section{The $\Lambda=0$ case.}
Its worth noticing that in some physically interesting situations the
reduction procedure eliminating higher order derivatives is redundant.
In the radiation filled universe with null cosmological constant the
correction $\hbar\alpha_1H^{(1)}_{ \mu\nu}$, which formally appears as
linear in $\hbar$, actually is quadratic, and consequently should be
abandoned as the $o(\hbar)$ term. To show this let us express the
traceless tensor $^{(1)}H^{\mu}_{\nu}$ in terms of the Ricci scalar and
the effective energy density

\[^{(1)}H=\frac R 2 (4\epsilon -R)\left[\begin{array}{cccc}
1&0&0&0\\
0&-1/3&0&0\\
0&0&-1/3&0\\
0&0&0&-1/3\end{array}
\right]\]

\noindent The field equations with the energy-momentum tensor
(\ref{Tef}) show that the scalar $R$ involves the trace of the tensor
$^{(3)}H^{\mu}_{\nu}$, namely $R=\alpha_3\hbar^{(3)}H^{\mu}_{\mu}$, so
it is a quantity linearly dependent on $\hbar$.

Writing $^{(3)}H^{\mu}_{\mu}$ in terms of the effective energy density
$\epsilon$ with the accuracy to terms $o(\hbar)$ we get $ R=\frac
43\alpha_3\hbar\epsilon^2$. Tensors $^{(1)}H^{\mu}_{\nu}$ and
$^{(3)}H^{\mu}_{\nu}$ can be rewritten as

\[^{(1)}H^{\mu}_{\nu}=\frac 83\alpha_3\hbar\epsilon^3\left[\begin{array}{cccc}
1&0&0&0\\
0&-1/3&0&0\\
0&0&-1/3&0\\
0&0&0&-1/3\end{array}
\right]\]
\[^{(3)}H^{\mu}_{\nu}=\frac 1 3 \epsilon^2\left[\begin{array}{cccc}
-1&0&0&0\\
0&5/3&0&0\\
0&0&5/3&0\\
0&0&0&5/3\end{array}
\right]-\frac 8{27}\alpha_3\hbar\epsilon^3\left[\begin{array}{cccc}
0&0&0&0\\
0&1&0&0\\
0&0&1&0\\
0&0&0&1\end{array}
\right]\]

\noindent Now it is clear that only the second of the expressions
$\alpha_1\hbar^{(1)} H^{\mu}_{\nu}$ and
$\alpha_3\hbar^{(3)}H^{\mu}_{\nu}$ is essentially linear in $\hbar$ and
forms the first-order quantum contribution to the energy-momentum
tensor. The first one $\alpha_1\hbar^{(1)} H^{\mu}_{\nu}$, which carries
all higher derivatives is actually square in $\hbar$. This is closely
related to the absence of particle creation in the radiation-filled
Robertson-Walker universe (see \cite{parker1983} and papers cited
there.) The theory with the energy-momentum tensor
$T_{\mu\nu}=T_{\mu\nu}^{(rad)}-\hbar\alpha_ 3H^{(3)}_{\mu\nu}$ leads to
the effective equation of state $P=\frac 13\epsilon -\frac
{4\alpha_3\hbar}9\epsilon^2$, which is perfectly consistent with
(\ref{algestate}).

\section{Stability of the empty space - dynamical systems approach.}

The equation of state of the form $P=P(\epsilon ,\Lambda )$ (or more
generally $P=P(\epsilon ,\Lambda ,H)$ see \cite{Khalat}) uniquely
determines cosmological evolution. The system (\ref{Ray}-\ref{kont}),
which defines the universe dynamics in the $\{H, \epsilon\}$-phase space
is autonomous. Choosing a point in the $\{H, \epsilon\}$-phase space,
one determines uniquely the metrics in the initial moment as well as the
metrics' evolution in time.\footnote { We abandon here a trivial freedom
to multiply the flat universe metrics by a factor constant in time.}
The stability of the Minkowski space-time is defined by the stability of
the $(H,\epsilon )=(0,0)$ point in the $\{H, \epsilon\}$-phase space
under the condition $\Lambda=0$. For the equation of state (\ref{algestate})
discussed in the preceding section the autonomous system (\ref{Ray}-\ref{kont})
reads:

\begin{eqnarray}
\frac {dH}{dt}&=&-H^2-\frac 1 3 \left[\epsilon -\frac 2 3 \alpha_3\hbar\epsilon^2\right]
\label{Hdot}\\
\frac {d\epsilon}{dt}&=&-4H\left[\epsilon -\frac 13\alpha_3\hbar\epsilon^2\right]
\label{edot}\end{eqnarray}

\noindent and its trajectories form levels of the integral
 
\begin{equation}H^2=\frac {\epsilon}3-K\sqrt {\frac {\epsilon}{G\epsilon_0a_0^4}}
-\frac {\alpha_3\hbar}6K\sqrt {\frac {\epsilon^3}{G\epsilon_0a_0^4}}
\label{calka}\end{equation}

\noindent The phase portrait of the system (\ref{Hdot}-\ref{edot})
is shown on Fig. 1. For completeness and also for readers convenience, we
attach Fig. 2. showing classical Friedmanian dynamics in the same
representation.  The phase structure of classical
radiation-filled universes and the phase structure defined by
(\ref{Hdot}-\ref{edot})  are topologically equivalent in the low energy
limit. This is so because one cannot enrich the structure of the $\{H,
\epsilon\}$-phase plane without violating the standard energy
conditions. On the other hand, according to (\ref{algestate}) these
conditions are well fulfilled for low and positive energy densities. The
equation of state (\ref{algestate}) formally admits violation of the
energy conditions but these states appears already in the Planckian
regime and hence, far beyond the region where semiclassical
approximation is valid. (The dotted region in the upper part of Fig. 1,
which contains three 'additional' critical points must be recognized as
nonphysical).

\hfill
\begin{minipage}{6.5cm}
\begin{figure}[h]
\centerline{\epsfig{file=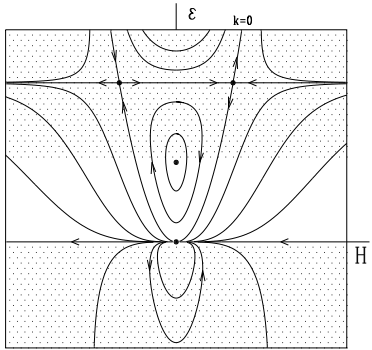, width=7cm}}
\caption{Dynamics of semiclassical radiation filled universes.}
\end{figure}
\end{minipage}
\hfill
\begin{minipage}{6.5cm}
\begin{figure}[h]
\centerline{\epsfig{file=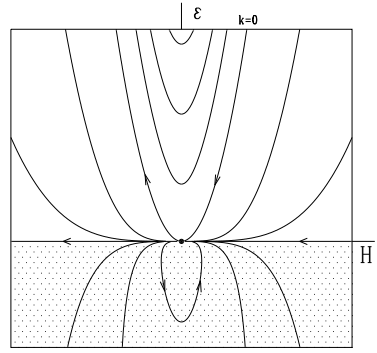, width=7cm}}
\caption{Dynamics of Friedmanian (classical) radiation filled universes.}
\end{figure}
\end{minipage}
\hfill
\vspace*{0.5cm}

An essential property of the system (\ref{Hdot}-\ref{edot}) is the absence
of solutions that change the energy density from positive to negative,
or the reverse. (Such behaviour was possible in the original semiclassical
theory and disqualified the empty space as a ground state.) Indeed, on
the strength of (\ref{algestate}), the initial condition  $\epsilon=0$
results in $\epsilon+P=0$, and consequently the right-hand side of
equation (\ref{edot}) vanishes. Both conditions $\epsilon =0$ and
$d\epsilon /dt=0$ ensure that the state of the zero energy density is
'persistent'. This is consistent with the results based on the
functional integral formalism \cite{mazur}, where all higher derivative
terms responsible for instability are eliminated by regularisation of
the energy-momentum tensor.

The stability of Minkowski space-time is the same as in the classical
theory. In both cases, the classical or the quantum, the
$(H,\epsilon)=(0,0)$ point is a three-fold point with elliptical sector
\cite{Andronov} and its type does not depend on the value of $\hbar$.
This means that the phase space is structuraly stable against quantum
corrections in the low energy density limit. This nontrivial property
does not follow from the solutions analicity in $\hbar$, but from the
form of the energy density tensor (\ref{Tef}) \footnote{In general, a
three-fold point may bifurcate into simple critical points under smooth
changes of the equation coefficients. This is what occurs when
cosmological constant appears. Solutions are analytical in $\Lambda$,
though the critical point corresponding to empty space bifurcates into
three simple points. Two of them represent de Sitter space time, the
third one -- the Einstein static universe \cite{Woszcz87}. However, no
bifurcation, results from quantum corrections.}.

\section{Summary and conclusions}

In the reduced Simon-Parker theory the energy-momentum tensor is
renormalized to take the hydrodynamic form with a simple, barotropic
equation of state.

The self-consistence conditions for semiclassical cosmology can be
imposed on observable quantities and weakened. By demanding the Hubble
expansion rate to be perturbative in $\hbar$ we allow the space
curvature to alter from 0 while quantum corrections to the flat
universes occur.

In the particular case of the radiation-filled universe and vanishing
cosmological constant, the dynamics of the Robertson-Walker universe in
the (original) semiclassical theory is described by a second order
equation, therefore it does not need either the reduction or
additional conditions of self-consistence. The reason lies in the
absence of particle creation in the radiation filled universe, which
manifests itself as an additional factor $\hbar$ 'hidden ' in the tensor
$^{(1)}H_{\mu\nu}$. This eventually eliminates all the higher derivative
terms.

Minkowski space-time has the same stability character as for
Einsteinian gravity, which is consistent with results based on the
functional integral formalism \cite{mazur}. The stability of Minkowski
space-time is independent of the numerical value of the Planck constant.
In the language of dynamical systems theory, this property is called
the structural stability of the $\{H,\epsilon\}$-phase space against
changes of $\hbar$.

Its worth noticing that the Liapunov stability of the environment with
equation of state (\ref{algestate}) with respect to
position-dependent perturbations is also the same as for the classical
radiation-filled universe\cite{SiemWo}, in contrast to the original
semiclassical theory, where quantum corrections let inhomogeneities
grow. This suggests an insignificant role for semiclassical corrections
in the processes of structure formation in the early universe.

\section*{Acknowledgements}
We would like to acknowledge Prof. Marek Demia\'nski and Prof. Lech
Soko{\l}owski for helpful discussion. This work was partially supported
by Polish research project KBN Nr 2 P03D 02210.

{}


\begin{thebibliography}{}
 \bibitem{Simon} J. Z. Simon, \Journal\PRD{43}{3308}{1991}
 \bibitem{PS93} L. Parker, J. Z. Simon, \Journal\PRD{47}{1339}{1993}.
 \bibitem{kwantinst}
G. T. Horowitz and R. M. Wald, \Journal\PRD{17}{414}{1978},\\
J. B. Hartle and G. T. Horowitz, \Journal\PRD{24}{257}{1981},\\
M. W. Suen, \Journal\PRD{40}{315}{1989}.
\bibitem{starobinski}A. Starobinski, \Journal\PLB{91}{99}{1980}.
\bibitem{Mazzitelli} F. D. Mazzitelli, \Journal\PRD{45}{2814}{1992}.
\bibitem{Simon2} J. Z. Simon, \Journal\PRD{45}{1953}{1992}.
\bibitem{Schmidt}H. J. Schmidt, \Journal\PRD{49}{6354}{1994},\\
                 A. B. Mayer and H. J. Schmidt, \Journal\CQG{10}{2441}{1993}.
\bibitem{Agu} I. M. Aguirregabiria, A. Hernandez and M. Rivas, \Journal\JPA{19}{L651}{1997}.
\bibitem{FHH} M. V. Fischetti, J. B. Hartle and B. L. Hu, \Journal\PRD{20}{1757}{1979},\\
                 L. H. Ford, \Journal\PRD{35}{2955}{1987},\\
                 B. L. Hu, \Journal\PLA{90}{375}{1982}.
\bibitem{SWAC} G. Siemieniec-Ozi\c{e}b{\l}o, A. Woszczyna,
              Acta Cosmologica, XXIII, 1, 1997.
\bibitem{parker1983} L. Parker, \Journal\PRL{50}{1009}{1983}.
\bibitem{Khalat} V. A. Belinsky, I. M. Khalatnikov, \Journal\JETP{42}{205}{1975},\\
V. A. Belinsky, I. M. Khalatnikov, \Journal\JETP{45}{1}{1977}.
\bibitem{mazur} P. O. Mazur, E. Mottola, \Journal\NPB{341}{187}{1990}.
\bibitem{Andronov} A. A. Andronov, E. A. Leontovich, I. I. Gordon and A.G. Maier,\\
{\em Qualitative Theory of Second Order Dynamic Systems.\\ \/} (Wiley, New York 1973).
\bibitem{Woszcz87} A. Woszczyna, \Journal\GRG{19}{233}{1987}.
\bibitem{SiemWo} G. Siemieniec-Ozi\c{e}b{\l}o, A. Woszczyna,\\
              "Scalar perturbations in open semiclassical FLRW universes." \\
              (to appear in Proceedings of The Eight
              Marcel Grossmann Meeting on General Relativity)
\end{thebibliography}
\end{document}